# CLOUD COMPUTING AND CONTENT MANAGEMENT SYSTEMS: A CASE STUDY IN MACEDONIAN EDUCATION

Jove Jankulovski[1] and Pece Mitrevski[2]

Faculty of Information and Communication Technologies,
University "St. Kliment Ohridski", Bitola, Republic of Macedonia

***ABSTRACT***

*Technologies have become inseparable of our lives, economy, and the society as a whole. For example, clouds provide numerous computing resources that can facilitate our lives, whereas the Content Management Systems (CMSs) can provide the right content for the right user. Thus, education must embrace these emerging technologies in order to prepare citizens for the 21st century. The research explored 'if' and 'how' Cloud Computing influences the application of CMSs, and 'if' and 'how' it fosters the usage of mobile technologies to access cloud resources. The analyses revealed that some of the respondents have sound experience in using clouds and in using CMSs. Nevertheless, it was evident that significant number of respondents have limited or no experience in cloud computing concepts, cloud security and CMSs, as well. Institutions of the system should update educational policies in order to enable education innovation, provide means and support, and continuously update/upgrade educational infrastructure.*

## 1. INTRODUCTION

The development of information and communication technologies – ICTs results into inevitable penetration of ICTs and Internet into every pore of our lives. At least once, every one of us did some actions online: checked web-based e-mail, changed image size online, edited A/V record, stored data, surveyed or took part in surveys, used various forms/templates, used software. All of this was done somewhere in the clouds without we possessing/using software locally or engaging local processing/computing resources. Clouds and cloud computing are available for a period of time. Each one of us used cloud resources, sometimes even without being aware of that, regardless of the services used: software, platform, infrastructure [1]. Research done by Version One in the UK, referred to on the blog of Load Storm in June 2009, reveals that at the early days of cloud computing: "41% of senior IT professionals (IT Directors/Managers in UK firms) admit that they "don't know" what cloud computing is"; and "66% of UK senior finance professionals (finance directors and managers) are confused about cloud computing" [2]. In such case, series of questions logically arise: Are we aware of the benefits and challenges of cloud computing? To what extent we are exploring possibilities that are available? Do we know enough about the possibilities? How much we used the possibilities? Are we aware of the challenges/ risks of cloud computing? What do we do to minimise challenges/risks? Do we have a backup plan to protect ourselves from challenges/risks that come with clouds?

Currently, the volume of information available on the Internet is continuously increasing as new content is published each second. Under these circumstances a number of challenges emerge: How do we maintain attractiveness of digital products and services in the infinite pool of digital content and information? How do we secure accurate and up to date content? How do we update web quickly and easily without losing links? Luckily, content management systems – CMSs are developed to facilitate our lives in this domain. CMSs support creation, management, distribution and publication of content, including corporate. They secure the right information for the right user; organise and provide access to all types of digital content; contain information about files and links to files for easy location and access. Such systems can encompass whole content created in an organisation.

Considering the above mentioned context, the education cannot function isolated from these trends and tendencies. It is clear that influence of globalisation and ICTs over education is enormous and has numerous and significant implications. Logically, the education should embrace these challenges and adjust its modus operandi in order to support building of digital citizenship. Therefore, the aim of this research was to examine ‘if’ and ‘how’ the application of cloud computing influences the application of CMSs and ‘if’ and ‘how’ it fosters usage of mobile devices to access cloud resources in the educational system in the Republic of Macedonia. In the remainder of this paper we discuss the elements of the research methodology in Section 2, present a selection of research findings in Section 3, and draw some conclusions and recommendations in Section 4.

## 2. Research Methodology Elements

The research methodology was constructed around the guiding/general hypothesis of the research: “Application of cloud computing concepts fosters application of content management systems in the educational system in Macedonia”. The design of the research was exploratory case study [3] in order to enable phenomena description in the current contextual setting. The primary research target was put on teachers in primary and secondary schools in Macedonia. Such population was heterogeneous from diverse aspects and contributes to the research comprehensiveness, as well as, to the research results. Limited number of students was encompassed as minor group with a secondary focus. The intention was to check if there will be a difference, and its potential extent, between considerations and perspective of various actors and age groups in the education related to the interest spheres of this research. From the population of all teachers and students in public primary and secondary schools, teachers and students in urban public primary and secondary schools were stratified [4]. In the framework of the research were scoped 178 teachers and students from urban schools in 10 cities throughout Republic of Macedonia. The sample was representative as it included geographically distributed schools; covered all types of public schools: primary schools, secondary vocational schools, and secondary high schools. The data gathering was done with a questionnaire that contained 12 close-ended questions and 8 open-ended questions. In the introductory section of the questionnaire were described key terms and notions, and some examples were provided, respectively. To this end, to further facilitate the process of filling in the questionnaire, there were illustrative examples for some of the open-ended questions. Like this, the questionnaire was easy to understand and was easy to fill in. It was prepared and published in Google Forms, easily accessible. Despite the possibility that Google Forms offer – compulsory response to each of the questions, this possibility was not used as it could have been expected that some of the respondents were not experienced enough in the target areas, thus might not have an opinion or might not feel confident to reply to some of the questions. At the same time, the intention was to compare how much received responses differ from provided examples in order to acquire an indication on experience of respondents in using clouds and content management systems. The link, accompanied with a request and directions, was distributed to sampled schools.

The data gathered via received responses on the survey undergone statistical analysis. All of the questions were analysed one by one. In addition, one of the analytical approaches applied was cross-tabulation. With this tool frequency of answers from two or more questions were combined in order to derive better insight of the views and perspectives of respondents regarding cloud computing and content management systems. To this end, selected findings are presented in this paper in order to illustrate the most specific considerations from this research.

## 3. Selected Research Findings

The gender, the age, and the school types of the respondents match the distribution of teachers across these criteria in all schools in the Republic of Macedonia. Therefore, the research scoping was well balanced and established solid grounds for valid and relevant research outcomes.
The broadband Internet coverage in Macedonia is more than 90% of the territory, and more than 95% of the public educational institutions. Also, Macedonia has the strongest IP backbone installed in the region. In addition, educational institutions are highly equipped with computers and laptops that practically each student can individually use a computer in school and each teacher can individually use laptop at work or at home. Still, during the research, the following finding surfaced: teachers prefer to fill in paper copy of the questionnaire instead of the digital version available online.

### 3.1. Analysis Of Selected Questions

The existing ICT infrastructure in the country is predominantly installed 10 years ago. Therefore, it was expected teachers to have long experience in using ICTs. The analysis confirmed this expectation. The received responses presented on the following figure show that respondents, use ICTs between 3 and 10 years (≈61% for professional purposes and ≈46% for private purposes), predominantly. Also, number of respondents who use ICTs more than 10 years is significant (≈22% for professional purposes and ≈34% for private purposes), especially for private purposes. *This finding provides solid research foundation as scoped respondents have considerably long ICT experience. Consequently, answers provide realistic picture of the state-of-art situation regarding the application of cloud computing concept and its potential influences on application of content management system in the educational system in Macedonia.*

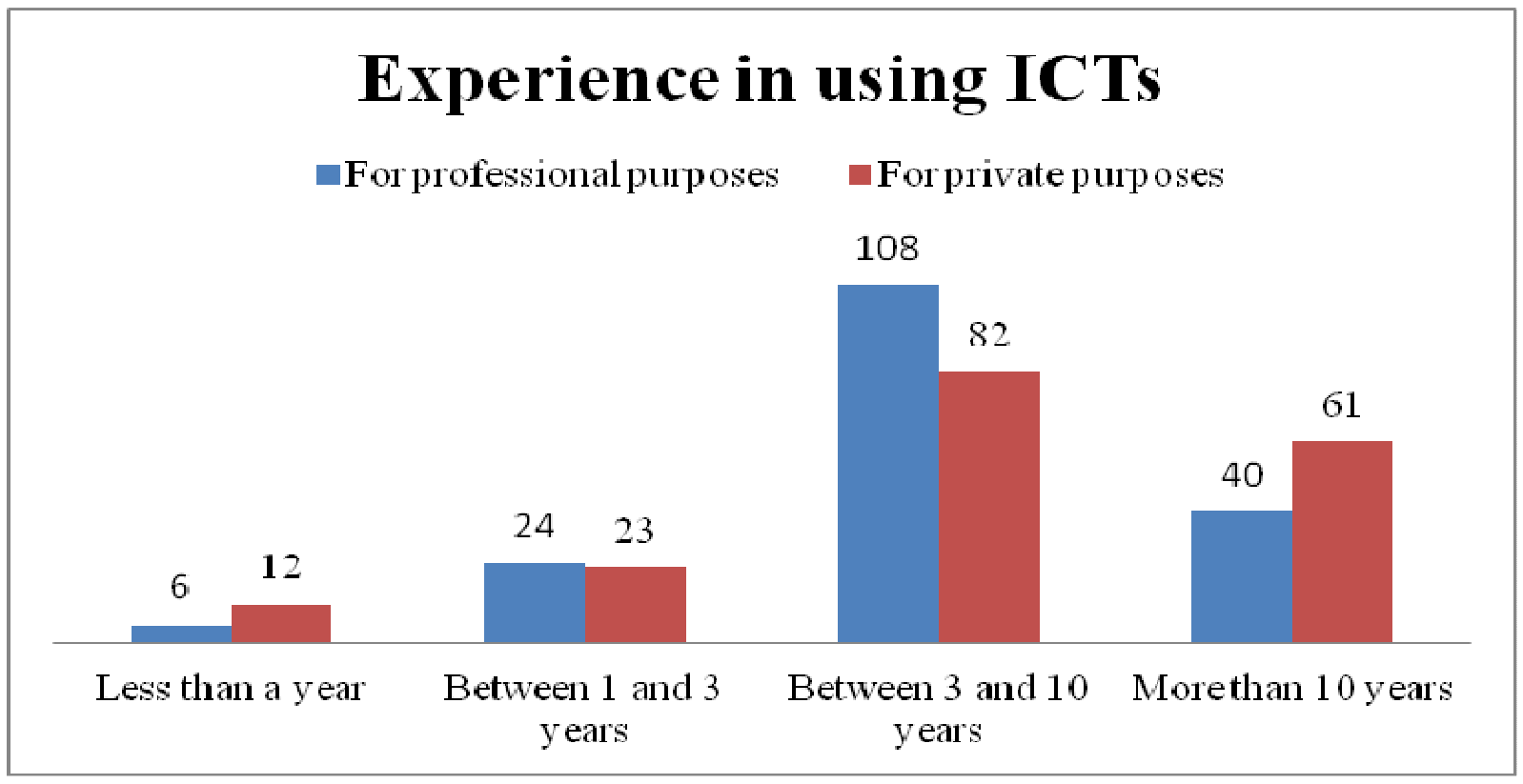

**Figure 1**. Distribution of respondents across years of experience in using ICTs: (a) professional purposes, and (b) private purposes

The research touched upon the experience of respondents in using clouds and CMSs. The received results are given on the Figure 2. It is apparent that respondents are more experienced in using clouds compared to CMSs. Also, from the below figure is clear that the respondents did not provide response (≈7% for clouds and ≈42% for CMS) and that number of respondents that have limited experience (less than a year is ≈12% for clouds and ≈22% for CMS) should not be neglected, especially when it comes to usage of CMSs. *This is indicative as it shows that although respondents are relatively experienced in using ICTs, they have some gaps or reservations in using clouds, and they lack experience or confidence in using CMSs.*

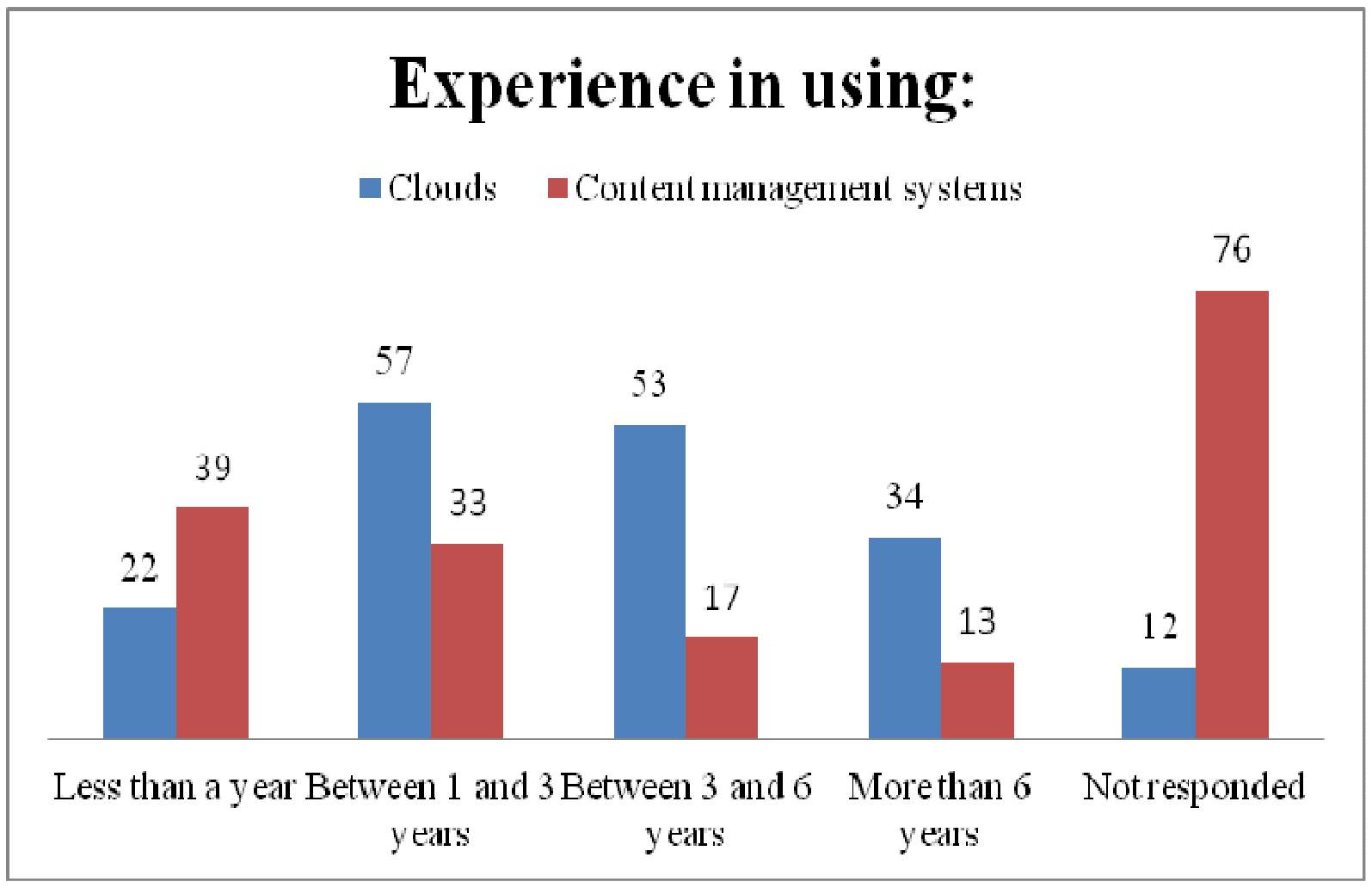

**Figure 2.** Distribution of respondents across years of experience in using: (a) clouds, and (b) content management systems – CMSs

The question regarding fostering effect of cloud services on application of CMSs revealed that *predominant percent of the respondents (42%) think that cloud services foster, moderately or significantly, application of CMSs*. From the Figure 3 it is also evident that 38% of the respondents did not respond, which is relatively high percentage. This denotes that *these respondents do not know whether cloud services influence the application of CMSs, or not, if at all.*

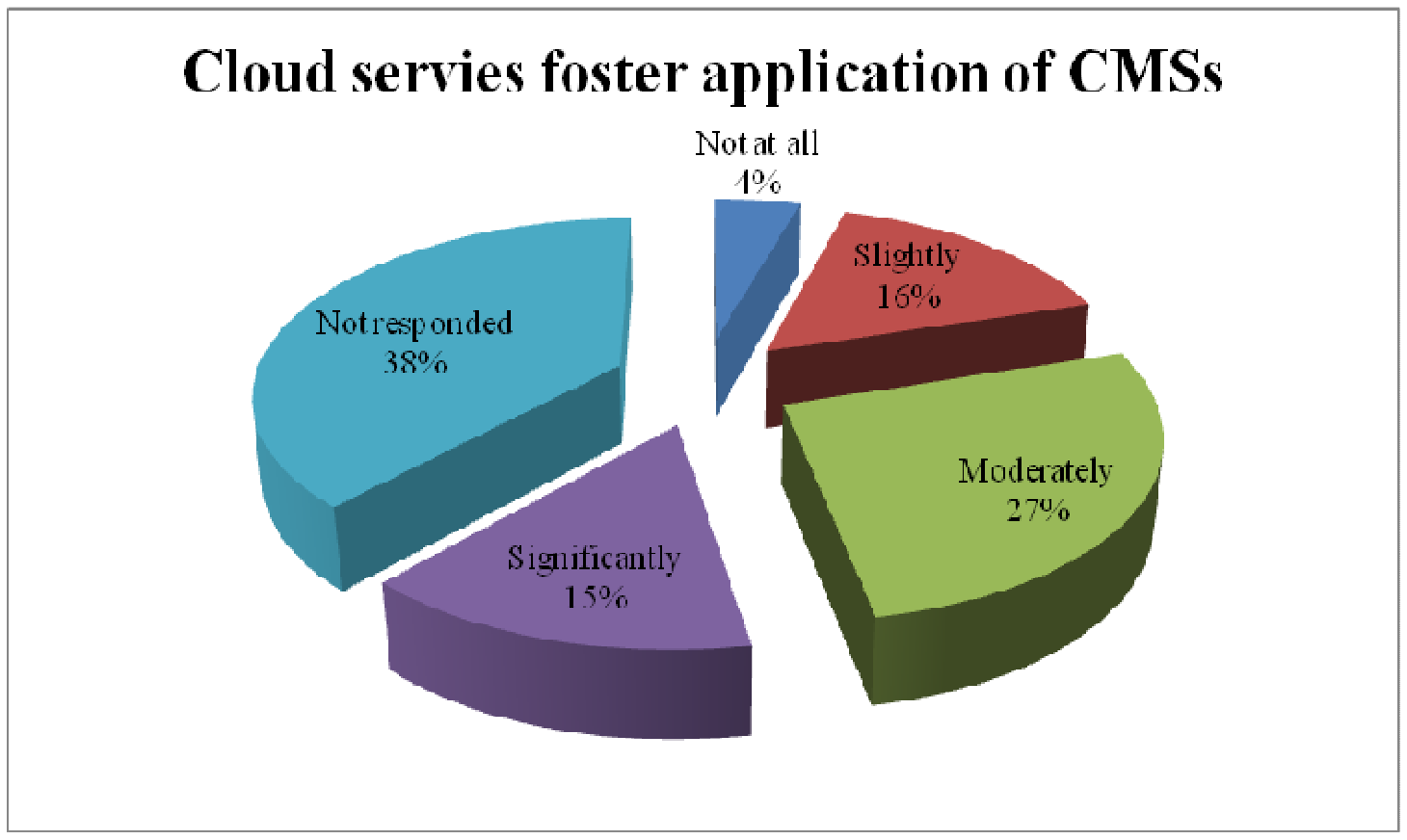

**Figure 3.** Cloud services foster intensity on application of CMSs

Awareness about the security of data and applications in the cloud was and is important issue that deserves attention, especially, due to the fact that security in non-cloud-based IT systems is much simpler and differs from the security in cloud-based IT systems [5]. Therefore, the respondents were asked about their opinion/awareness regarding the security of data transferred to a cloud. It is interesting to observe that 38% of the respondents did not have an opinion, 7% did not respond, and 3% tend not to use cloud for data storage. All of these responses *present misconception, or incomplete awareness among respondents about security of data in a cloud.*

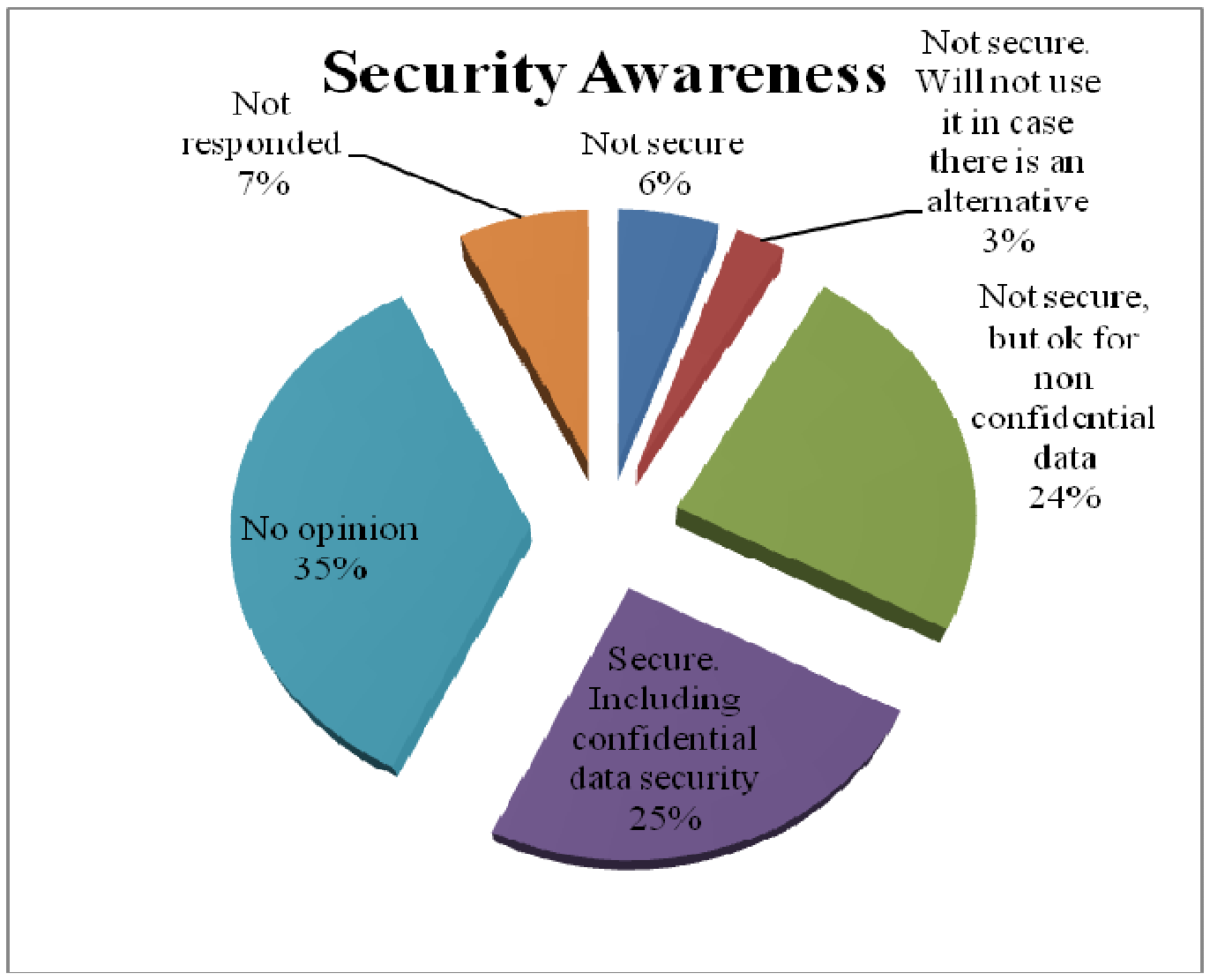

**Figure 4**. Awareness regarding cloud data security

Smart phones, tablets, and other handheld devices are inseparable part of our everyday work, daily routines, life. They are surely taking over our private time; becoming all kinds of libraries that we want/need; are handy personal and professional assistants and/or learning devices; even are repositories of our digitalised life and personal/private data. Hence, it can be considered that we use these kinds of devices for multiple purposes at arbitrary moments in time. Considering this, the research checked the-cloud services facilitation extent on the usage of mobile / handheld devices in comparison to stationary devices. It can be noted that dominant percent of respondents think that extent of facilitation is moderate (38%) and significant (30%). Still, the other alternatives (no response, not, slightly) should be considered, as they represent considerable part of the answers pool. *This shows that 32% of the respondents are not completely sure about the cloud computing concept or how cloud services facilitate work of applications on handheld devices.*

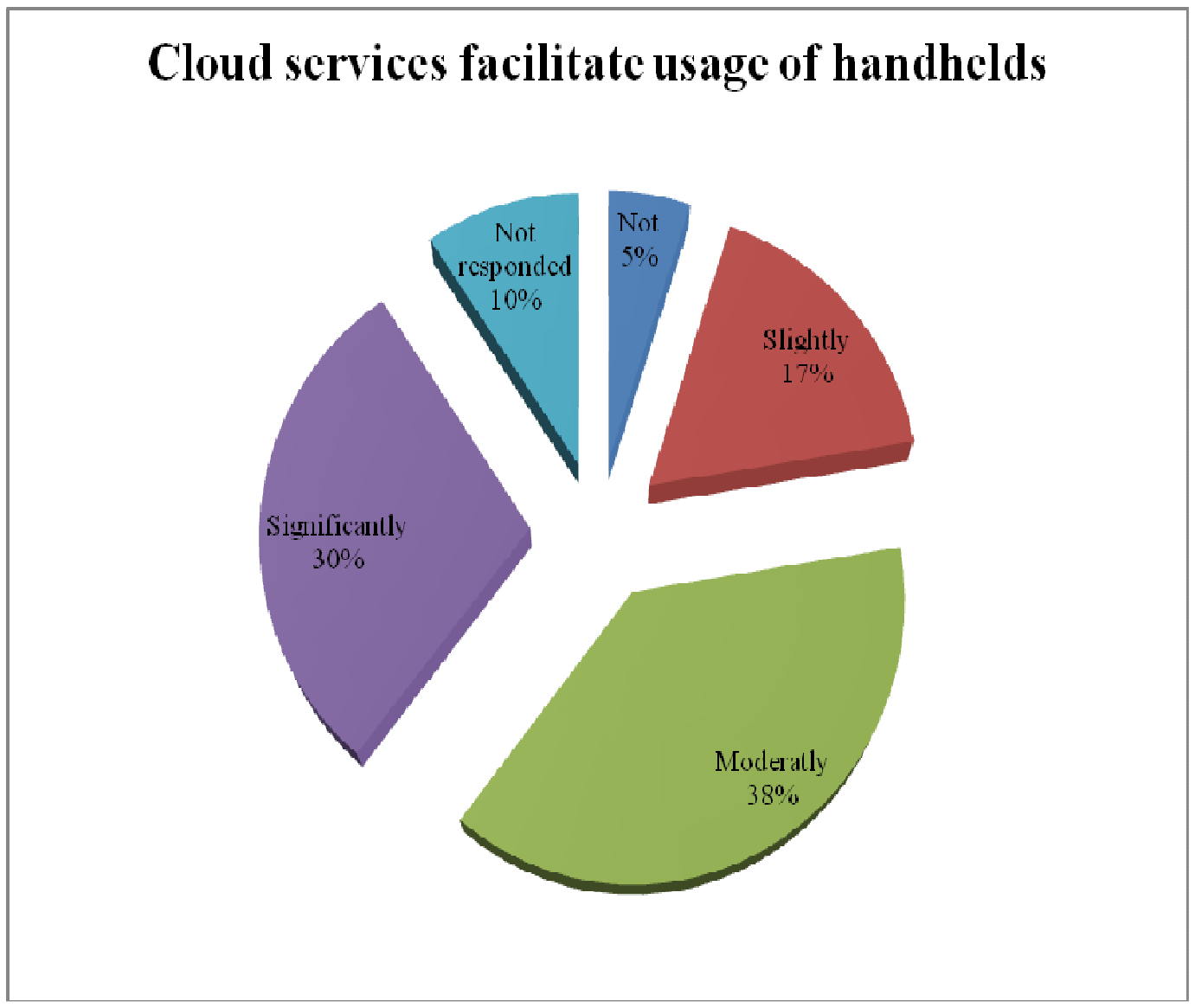

**Figure 5.** Cloud services facilitate usage of handheld devices

### 3.2. ANALYSIS OF SELECTED CROSS-TABULATED QUESTIONS

Cross-tabulation of questions revealed a number of eye-opening findings. One of them is cross-tabulation of years of experience in using ICTs with awareness of respondents regarding data security in a cloud. The data presented in the Table 1 *show that respondents' confidence in security of data in a cloud declines with years of experience in using ICTs. Or, the more teachers are ICT experienced the more they are reserved regarding the data security.* This is an interesting message to cloud providers as it says that experienced ICT users encountered glitches in their work with clouds, so they developed some reservations regarding data security
.

**Table 1.** Cross-tabulation of experience in using ICTs and cloud data security awareness.

| % | **Not secure** | **Not secure. Will not use it in case there is an alternative** | **Not secure. Ok for non confidential data** | **Secure, including confidential data security** | **No opinion** | **Total** |
|---|---|---|---|---|---|---|
| < 1 year | 16.67 | 16.67 | 33.33 | 16.67 | 16.67 | 100.00 |
| 1 – 3 years | 0.00 | 4.76 | 14.29 | **33.33** | 47.62 | 100.00 |
| 3 – 10 years | 7.07 | 2.02 | 21.21 | **30.30** | 39.39 | 100.00 |
| > 10 years | 5.13 | 2.56 | **43.59** | 15.38 | 33.33 | 100.00 |

The next finding was derived from cross-tabulation of received answers on awareness regarding the security of data in a cloud across the age of the respondents. The finding derived from the Table 2 speaks similarly as the previous finding – *the confidence in data in a cloud is higher among younger respondents compared to older ones – teachers.* This confirms the finding that surfaced with previous cross-tabulation. This is one of the examples proving that introduction of minor group with secondary focus was relevant and valuable.

**Table 2.** Cross-tabulation of age of the respondents and cloud data security awareness.

| % | **Not secure** | **Not secure. Will not use it in case there is an alternative** | **Not secure. Ok for non confidential data** | **Secure, including confidential data security** | **No opinion** | **Total** |
|---|---|---|---|---|---|---|
| 14 – 18 years | 3.03 | 6.06 | 15.15 | **60.61** | 15.15 | 100 |
| 19 – 25 years | 0.00 | 12.50 | **25.00** | 18.75 | 43.75 | 100 |
| 26 – 49 years | 3.53 | 1.18 | **29.41** | 23.53 | 42.35 | 100 |
| > 50 years | 19.35 | 0.00 | **29.03** | 3.23 | 48.39 | 100 |

The received answers on question exploring the cloud services facilitation extent on the usage of handheld devices in comparison to stationary devices were cross-tabulated with the gender, age and experience in using ICTs. The table 3 informs us that genders' opinions are similar (male ≈43% and female 42% respondents replied moderately) – *cloud services are moderately facilitating the usage of mobile / handheld devices.* Or, cloud services are not determining the usage of mobile / handheld devices, but only facilitating the usage.

**Table 3**. Cross-tabulation of gender of the respondents and cloud services facilitation extent on usage of mobile / handheld devices.

| % | Not | Slightly | Moderately | Significantly | Total |
|---|---|---|---|---|---|
| Male | 7.41 | 11.11 | **42.59** | 38.89 | 100.00 |
| Female | 4.67 | 23.36 | **42.06** | 29.91 | 100.00 |

The data in Table 4 reveal that younger respondents, students and younger teachers think that cloud services moderately facilitate usage of mobile / handheld devices. Namely, 70% of the respondents on age 14 – 18 years, responded moderately and 69% of the respondents on age 19 – 25 years responded moderately, too. While, more senior teachers think that cloud services significantly facilitate usage of mobile / handheld devices (26 – 49 years of age, 40% responded significantly and older than 50 years 43% responded significantly). *Or, with age, respondents increasingly think that cloud services facilitate usage of mobile / handheld devices*.

**Table 4**. Cross-tabulation of age of the respondents and cloud services facilitation extent on usage of mobile / handheld devices.

| % | Not | Slightly | Moderately | Significantly | Total |
|---|---|---|---|---|---|
| 14 – 18 years | 10.00 | 10.00 | **70.00** | 10.00 | 100.00 |
| 19 – 25 years | 0.00 | 12.50 | **68.75** | 18.75 | 100.00 |
| 26 – 49 years | 5.88 | 23.53 | 30.59 | **40.00** | 100.00 |
| > 50 years | 3.33 | 20.00 | 33.33 | **43.33** | 100.00 |

The analysis of the data cross-tabulated in the Table 5 shows that, in principle, *the opinion of the respondents on cloud services facilitation extent on usage of mobile / handheld devices improves with their experience in using ICTs.* There is small difference in this consideration among the respondents in the group of 3 – 10 years of experience in using ICTs.

**Table 5.** Cross-tabulation of experience in using ICTs and cloud services facilitation extent on usage of mobile / handheld devices.

| % | Not | Slightly | Moderately | Significantly | Total |
|---|---|---|---|---|---|
| < 1 year | 33.33 | 16.67 | **33.33** | 16.67 | 100.00 |
| 1 – 3 years | 5.26 | 5.26 | **52.63** | 36.84 | 100.00 |
| 3 – 10 years | 5.00 | 27.00 | **44.00** | 24.00 | 100.00 |
| > 10 years | 2.78 | 5.56 | 33.33 | **58.33** | 100.00 |

The duration of usage of cloud services for professional purposes (in education), if at all, is presented on Table 6. The most prominent number of 41% (40.91%) is for *respondents that used ICTs in education for 1 to 3 years used cloud services in the same period.* The more *experienced respondents (3 – 10 years or more than 10 years of ICT experience in education), used cloud services predominantly 3 – 6 years and used even them less before 6 years.*

**Table 6.** Cross-tabulation of experience in using ICTs and experience in using cloud services.

| % | < 1 year | 1 – 3 years | 3 – 6 years | > 6 years | Total |
|---|---|---|---|---|---|
| < 1 year | 20.00 | 40.00 | 40.00 | 0.00 | 100.00 |
| 1 – 3 years | 31.82 | **40.91** | 18.18 | 9.09 | 100.00 |
| 3 – 10 years | 7.84 | 33.33 | 33.33 | 25.49 | 100.00 |
| > 10 years | 16.22 | 32.43 | 35.14 | 16.22 | 100.00 |

In the primary schools, the respondents used CMSs for period up to 3 years (28.13% less than a year and 31.25% for 1 – 3 years). The percent of respondents who used CMSs, in period up to 3 years, in the high schools are higher – 43.33%, for the same periods, respectively. While in secondary vocational schools CMSs are predominantly used in the last year and at the same time CMSs were used in a more balanced way over longer period of time (more than 6 years). *Experience in using CMSs varies with the different type of schools along same parameters: time and intensity.*

**Table 7**. Cross-tabulation of school type and experience in using CMSs

.

| % | < 1 year | 1 – 3 years | 3 – 6 years | > 6 years | Total |
|---|---|---|---|---|---|
| Primary school | **28.13** | **31.25** | 21.88 | 18.75 | 100 |
| Secondary high school | **43.33** | **43.33** | 6.67 | 6.67 | 100 |
| Secondary vocational school | **42.50** | 25.00 | 20.00 | 12.50 | 100 |

*The respondents who are less experienced in ICTs used CMSs in a shorter period of time*, or as Table 8 shows, 1 – 3 years of ICT experience results into 75% of respondents using CMSs less than a year. *In principle, experience in using ICTs is directly linked with experience in using CMSs.* The only exception to this finding is the case of respondents who use ICTs for more than 10 years to use CMSs less than a year. Only 1 respondent used ICTs up to a 1 year used CMSs up to a year, too.

**Table 8**. Cross-tabulation of experience in using ICTs and experience in using CMSs.

| % | < 1 year | 1 – 3 years | 3 – 6 years | > 6 years | Total |
|---|---|---|---|---|---|
| < 1 year | 100.00 | 0.00 | 0.00 | 0.00 | 100.00 |
| 1 – 3 years | 75.00 | 25.00 | 0.00 | 0.00 | 100.00 |
| 3 – 10 years | 29.09 | 40.00 | 23.64 | 7.27 | 100.00 |
| > 10 years | 38.24 | 23.53 | 11.76 | 26.47 | 100.00 |

The Table 9 encompasses cross-tabulation of age of the respondents and opinion on how much could services might foster application of CMSs. It is apparent that *regardless of the age, predominant number of the respondents thinks that cloud services moderately foster application of CMSs.*

**Table 9**. Cross-tabulation of age of the respondents and cloud services fostering effect on application of CMSs.

| % | Not | Slightly | Moderately | Significantly | Total |
|---|---|---|---|---|---|
| 14 – 18 years | 0.00 | 26.67 | **40.00** | 33.33 | 100.00 |
| 19 – 25 years | 16.67 | 33.33 | **50.00** | 0.00 | 100.00 |
| 26 – 49 years | 7.81 | 21.88 | **42.19** | 28.13 | 100.00 |
| > 50 years | 8.00 | 32.00 | **48.00** | 12.00 | 100.00 |

The respondents with ICT experience of 1 – 3 years predominantly think that cloud services moderately (37.5%) and significantly (37.5%) foster application of CMSs, see Table 10. While the respondents with ICT experience of 3 – 10 years and more than 10 years think that cloud services moderately (44.07% and 44.45%, respectively) foster application of CMSs. By all means, it is evident that *significant percent of respondents think that cloud services do foster application of CMSs*.

**Table 10.** Cross-tabulation of experience in using ICTs and cloud services fostering effect on application of CMSs.

| % | Not | Slightly | Moderately | Significantly | Total |
|---|---|---|---|---|---|
| < 1 year | 0.00 | 50.00 | 50.00 | 0.00 | 100.00 |
| 1 – 3 years | 6.25 | 18.75 | **37.50** | **37.50** | 100.00 |
| 3 – 10 years | 3.39 | 32.20 | **44.07** | 20.34 | 100.00 |
| > 10 years | 15.15 | 15.15 | **45.45** | 24.24 | 100.00 |

## 4. Conclusions and Recommendations

Development of ICTs and internationalisation set new development roads for economies and societies. These development trends are changing the notion of citizenship from traditional into in digital citizenship. As a result, a shift in the requirements of skills and competencies for digital citizens is necessary. These new skills and competences are frequently called 21st Century Skills and Competences as these are features of the contemporary societies, only. There are numerous attempts to identify them. Here are listed several initiatives/documents: Cisco-Intel-Microsoft project for Assessment and Teaching for 21st Century Skills [6]; Framework for 21st Century Learning [7] by the organisation Partnership for 21st Century Learning; Organisation for Economic Cooperation and Development – OECD introduced, in some of the OECD member countries, Framework for Conceptualising of the 21st Century Skills and Competencies [8]; European Commission and its Key Competences for Lifelong Learning [9]. Such trends strongly demand from the educational systems to "equip" people, especially young, with new skills and competencies that will enable them to exercise their digital citizenship and to contribute to life in the knowledge based societies. Inevitably, these skills and competencies should be considered in all phases and elements of the teaching/learning process, in preparation of future teachers, and while using ICTs for arbitrary purposes.

Predominant number of respondents has experience in using clouds for a reasonably long time (up to 6 years). The awareness of the respondents regarding data security in clouds speaks of reservations in some cases when confidential data are considered and in fewer cases for non confidential data. In principle, not negligible number of "no responses" or "no opinions" was encountered. This means that significant number of respondents do not have enough experience, or no experience at all, in cloud computing as a concept and are not completely aware about the cloud security. The research revealed that clouds are necessary for better utilisation of the mobile technologies possibilities/resources.

The number of respondent that use CMSs is relatively low. Still, the respondents who use CMSs think that cloud services moderately or significantly facilitate the application of CMSs. Here, it is important to mention that respondents used CMSs only for some or limited purposes. Namely, respondents referred only to web development as application of CMSs, which is only one of the multiple purposes CMSs can be used for [10].

The analysis, quantitative and qualitative, showed that the respondents have tendency of not answering few of the close-ended, as well as, some of the open-ended questions. This is indicative and informs us that there areas targeted by those questions are not known enough to respondents, or that they do not have experience in the areas and therefore, or that the respondents are not feeling confident to answer.

There are teachers who have some experience in using clouds, CMSs and clouds services for mobile technologies. Yet, these teachers cannot be expected to implement the changes on a systemic level. In this regard, OECD research suggests that it cannot be expected students to achieve better in countries that only invested in introduction of ICTs in education [11].

Therefore, logically, this situation requires structured and systemic approach by the institutions of the system. It is evident that there exists urgent need, creators of the educational policies and actors in education to be acquainted with these concepts/systems, the possibilities and risks of their application, as much as possible, and as early as possible. The institutions of the system, in broad terms: should update educational policies in order to enable education innovation; develop means/tools for teachers and students to use during learning; encourage teachers and students to use these means/tools; create pleasant learning environment; provide permanent support to teachers and students; continuously update / upgrade created infrastructure.

More specifically, preparation of an ICT infrastructure, preferably CMS based, which directly will support application of cloud computing, mobile technologies and CMSs in education is necessary. Or, at minimum, preparation of materials for teachers that will provide them with inputs which technologies can be used for which purpose in education; and how to use such technologies. In parallel, there exists a need to initiate professional development of teachers (in-service training) that will target these technologies / concepts. An indicative but not exhaustive and final list of topics for professional development of teachers can be: 21$^{st}$ century skills and competences, intellectual property rights, cloud computing, using mobile technologies for learning, purposeful application of CMSs. At the same time, the curricula of higher education institutions should address these issues during pre-service training of future teachers. It is necessary to update curricula, thus, application of cloud services, mobile technologies, and CMSs is encouraged and supported. Like this, teachers' initiative will be boosted so teachers and students will become agents of educational innovation in teaching / learning process.

## AUTHORS

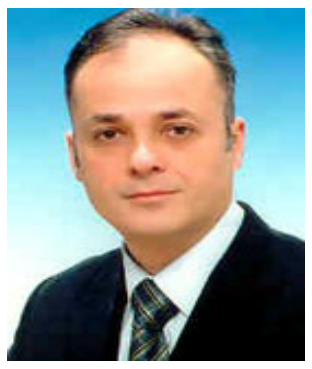

**Jove Jankulovski** is a PhD student at the Faculty of Information and Communication Technologies, University "St. Kliment Ohridski" – Bitola, Republic of Macedonia. He has long and comprehensive experience in teaching students on ICTs, computer programming and related courses. In parallel, he has extensive experience in training teachers on integration of ICTs in the curriculum. He set up online project based learning infrastructure in schools in Macedonia. As a result, he took part in numerous projects using ICTs. His research interests are e-Learning and immersive technologies.

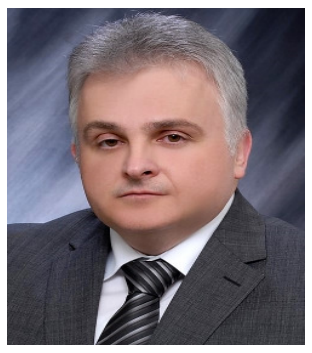

**Pece Mitrevski** received his BSc and MSc degrees in Electrical Engineering and Computer Science, and the PhD degree in Computer Science from the Ss. Cyril and Methodius University in Skopje, Republic of Macedonia. He is currently a full professor and Dean of the Faculty of Information and Communication Technologies, University "St. Kliment Ohridski" – Bitola, Republic of Macedonia. His research interests include Computer Architecture, Computer Networks, Performance and Reliability Analysis of Computer Systems, e-Commerce, e-Government and e-Learning. He has published more than 100 papers in journals and refereed conference proceedings and lectured extensively on these topics. He is a member of the IEEE Computer Society and the ACM.

## ANNEX – SURVEY QUESTIONS

1. **Gender** ☐ male ☐ Female
2. **Age** ☐ 14 – 18 ☐ 18 - 25 ☐ 26 - 49 ☐ more than 50
3. **City** _____________ Enter city name
4. **What is the type of school you work in or study?**
   ☐ primary school ☐ high school ☐ secondary vocational schools
5. **How many years you have used ICTs related to your work or studying**?
   ☐ less than a year ☐ 1 – 3 years ☐ 3 – 10 years ☐ more than 10 years
6. **How many years you have used ICTs for purposes that are not related to your work or study**?
   ☐ less than a year ☐ 1 – 3 years ☐ 3 – 10 years ☐ more than 10 years
7. **List *clouds* or *clouds services* that you have used or are using?** (Examples of clouds and clouds services: Google Drive, Google Cloud, Sound Cloud, Dropbox, Prezi, Microsoft Cloud, HTC Cloud, iCloud, Samsung Cloud, CloudZilla, etc.)
   __________________________________________________________________________
8. **List purposes for which you have used or are using *clouds*?**
   __________________________________________________________________________
9. **How many years have you used clouds? (in total for all purposes for which you have used or are using clouds)**
   ☐ less than a year ☐ 1 – 3 years ☐ 3 – 6 years ☐ more than 6 years
10. **To what extent are the data transferred to *the cloud* secure?**
    a) Not secure
    b) Not secure and if I have an option, I will not use a cloud
    c) Not secure, but for non confidential data using cloud is ok
    d) Secure for data of confidential nature
    e) Have no opinion regarding this question
11. **To what extent cloud services facilitate the usage of mobile / handheld devices** (for example, mobile phones, tablets, palm computers and other similar devices) **to access a cloud compared to the stationary devices?**
    a) Does not facilitate
    b) Slightly facilitates
    c) Moderately facilitates
    d) Significantly facilitates
12. **Describe your thinking regarding your answer to the previous question:**
    __________________________________________________________________________
13. **Rank the necessity of clouds for better facilitation of the mobile devices possibilities** (for example phones, tablets, and other similar devices). **1 is the lowest, 10 is the highest.**

| | | | | | | | | | |
|---|---|---|---|---|---|---|---|---|---|
| **1** | **2** | **3** | **4** | **5** | **6** | **7** | **8** | **9** | **10** |

14. **For which purpose/application potential clouds in the Republic of Macedonia can be introduced** (for example: education, e-commerce, information and document needs of citizens)**?** ____________________________________________________________
15. **List *content management systems* that you used or are using** (examples: WordPress, Joomla, Drupal, Plone, etc.)**?** _______________________________________________

16. **List the purposes for which you have used or are using *content management systems*?** ___________________________________________________________________

17. **For how many years in total you had used *content management systems*** (in total for all of the purposes for which you had used or are using content management systems)**?**
☐ less than a year ☐ 1 – 3 years ☐ 3 – 6 years ☐ more than 6 years

18. **To what extent *cloud service* fosters application of content management systems?**
    a) Does not foster
    b) Slightly fosters
    c) Moderately fosters
    d) Significantly fosters

19. **Describe your thinking regarding your answer to the previous question:** ___________________________________________________________________

20. **For which purpose/application can *content management systems* be used the most in the Republic of Macedonia? (Multiple responses are possible)**
    a) For the needs of instruction implementation in the educational system (primary, secondary, high)
    b) For the needs of preparation of teaching materials and teaching means in the educational system
    c) For following (controlling) the work of the personnel
    d) For permanent access to existing database, or creation of new databases
    e) As a platform for collaboration and/or for communication

Other ___________________________________________________________________